\documentclass[pre,twocolumn,aps,showpacs,showkeys]{revtex4}
\usepackage{graphicx}
\begin{document}

\markboth{G.P. TSIRONIS}{NONLINEAR MAGNETOINDUCTIVE TRANSMISSION LINES}

\title{NONLINEAR MAGNETOINDUCTIVE TRANSMISSION LINES 
}

\author{NIKOS LAZARIDES$^\dagger$, VASSILIS PALTOGLOU$^\ddagger$ and 
       G. P. TSIRONIS$^\star$
}
\affiliation{Department of Physics, University of Crete, and FORTH \\
P.O. Box 2208, 710 03, Heraklion, Crete, Greece\\
$^\dagger$nl@physics.uoc.gr \\
$^\ddagger$vaspal@physics.uoc.gr \\ 
$^\star$gts@physics.uoc.gr 
}

\begin{abstract}
Power transmission in one-dimensional nonlinear magnetic metamaterials
driven at one end is investigated numerically and analytically
in a wide frequency range. The nonlinear magnetic
metamaterials are composed of varactor-loaded split-ring resonators which are
coupled magnetically through their mutual inductances, forming thus a magnetoiductive 
transmisison line.    
In the linear limit, significant power transmission along the array only appears
for frequencies inside the linear magnetoinductive wave band.
We present analytical, closed form solutions for the magnetoinductive waves 
transmitting the power in this regime, and their discrete frequency dispesion.
When nonlinearity is important, more frequency bands with singificant power transmission
along the array may appear. In the equivalent circuit picture, the nonlinear 
magnetoiductive transmission line driven at one end by a relatively weak electromotive
force, can be modeled by coupled 
resistive-inductive-capacitive (RLC) circuits with voltage-dependent capacitance.
Extended numerical simulations reveal that power transmisison along the array is also
possible in other than the linear frequency bands, which are located close to the 
nonlinear resonances of a single nonlinear RLC circuit.
Moreover, the effectiveness of power transmission  for driving frequencies in
the nonlinear bands is comparable to that in the linear band. 
Power transmission in the nonlinear bands occurs through the linear modes of the system,
and it is closely related to the instability of a mode that is localized at the driven
site.
\end{abstract}


\maketitle

\section{Introduction}
\noindent 
The \emph{metamaterial} concept usually refers to a periodic arrangement of
artificially structured elements, designed to achieve advantageous and/or unusual
electromagnetic properties compared to those of naturally occuring materials
\cite{Smith2004}.
For example, the first metamaterial ever realized \cite{Shelby2001},
exhibited negative refraction index in a narrow frequency band around 5 Gigahertz.
The negative refraction index property requires that both the dielectric permittivity 
and the magnetic permeability are simultaneously negative. The most widely used 
elements for constructing a negative refractive index metamaterial are the 
electrically small resonant 'particles' called
\emph{split-ring resonators} (SRRs) and metallic wires. 
While the array of wires gives a negative permittivity below the plasma frequency,
the array of the SRRs gives a negative magnetic permeability
above their resonance frequency. 
Thus, the refraction index can be negative in a narrow frequency band.
Since their discovery, metamaterials have been the target of intensive 
study. Although the developement of metamaterials at microwave frequencies has 
progressed to the point where scientists and engineers are now pursuing 
applications, research on metamaterials that operate at higher frequencies
is still in an early stage \cite{Linden2006,Soukoulis2007,Litchinitser2008}. 

There is a wide subclass of metamaterials, that exhibit significant magnetic 
properties and even negative magnetic permeability at Terahertz and optical 
frequencies \cite{Yen2004,Katsarakis2005,Linden2006}, even though they are made 
of non-magnetic materials. Magnetism at such high 
frequencies is particularly important for the implementation of devices such as 
tunable lenses, adaptive mirrors, etc., since there are only a few natural 
materials that respond magnetically above microwaves. Moreover, magnetism in those
natural materials is usually weak and within narrow frequency bands,
limiting thus their use in possible Terahertz devices.
The most common realization of such a \emph{magnetic metamaterial} is composed
of periodically arranged SRRs in a one- or two-dimentional lattices.
In its simplest version an SRR is just a highly conducting 
\cite{Linden2004,Enkrich2005,Katsarakis2005} or superconducting \cite{Gu2010}
ring with a slit. The SRR structure was originally proposed by Pendry \cite{Pendry1999},
who theoretically predicted that a periodic SRR-based structure could exhibit a negative
magnetic permeability frequency band when irradiated by an alternating electromagnetic 
field of appropriate polarization. Pendry also realized that the SRR structure 
has considerable potential for enchancing nonlinear phenomena.
When the SRR dimensions are much smaller than the wavelength of the incident field,
the SRR can be regarded as an electric circuit consisting of inductive, capacitive 
and ohmic elements connected in series. The ring forms the inductance $L$, while the 
slit can be considered as a capacitor of capacitance $C$. 
The ohmic element $R$ may represent all possible losses of the structure.
 
The next step was naturally to built nonlinearity in the SRR structure.
Nonlinearity implies real-time tunability and multistability that is certainly 
a desired property of possible future devices. The SRRs can become nonlinear either by 
the insertion of a strongly nonlinear dielectric \cite{Hand2008} or a nonlinear 
electronic component (i.e., a varactor) \cite{Shadrivov2006a,Powell2007,Wang2008} 
into their slits. 
Both ways would lead to effectively field dependent magnetic
permeability for an array of nonlinear SRRs.  
Recently, the dynamic tunability by an external field for a two-dimensional array
of varactor-loaded SRRs was demonstrated experimentally \cite{Shadrivov2008}. 
The SRRs in such an array are coupled through magnetic and/or electric dipole-dipole 
interactions,
whose strength depends on the relative orientation of the SRRs and their slits in the
array \cite{Sydoruk2006,Hesmer2007,Penciu2008,Feth2010}. 

In the equivalent circuit picture, the dynamics of a periodic nonlinear SRR array
irradiated by an electromagnetic field, can be described by a set of coupled and driven
ordinary differential equations with on-site nonlinearity. The discreteness, which is 
inherent in those SRR-based magnetic metamaterials, along with the nonlinearity and the 
weak coupling between their elements,
allow for the generation of nonlinear excitations in the form of intrinsic localized 
modes or \emph{discrete breathers} \cite{Sievers1988}. 
These nonlinear modes appear generically in discrete and nonlinear extended systems.
Recent theoretical work in one- and two-dimensional nonlinear SRR lattices have 
demonstrated the 
existence and the stability of discrete breathers both in energy-conserving and 
dissipative 
systems \cite{Lazarides2006,Eleftheriou2008,Lazarides2008a,Eleftheriou2009}.
It has also demonstrated that breathers may be formed spontaneously through
modulational instability in binary nonlinear magnetic metamaterials 
\cite{Molina2009,Lazarides2009,Lazarides2010b}.
Moreover, domain-wall excitations \cite{Shadrivov2006b} and envelope solitons
\cite{Kourakis2007,Cui2009,Tsironis2010,Cui2010} 
may as well be excited in those systems, which 
seem to be stable even in the presence of noise \cite{Tsurumi2008}.

The SRR-based magnetic metamaterials support a new kind of electromagnetic waves, 
the magnetoinductive
waves, which exhibit phonon-like dispersion curves, and they can transfer energy
along the array \cite{Shamonina2004,Syms2006,Syms2010}. 
It is thus possible to fabricate a contact-free data and 
power transfer device, a {\em magnetoinductive transmission line},
which make use of the unique properties of the magnetic 
metamaterial structure, and may function as a frequency-selective communication
channel for devices via their magneto-inductive wave modes \cite{Stevens2010}.    
In the present work we investigate the power transmission along a one-dimensional 
varactor-loaded SRR-based magnetic metamaterial, in the two possible geometries,
which is driven at one end by a sinusoidal power sourse.
In the linear limit, energy tranfer along the array occurs only for driving 
frequencies in the linear magnetoinductive wave band. However, the nonlinearity 
could generate more frequency bands where efficient power transmission along the 
array is possible. Power transmission in chains of coupled anharmonic oscillators
for driving frequencies in the band gap of the linear spectrum has been recently 
investigated, and that effect is referred to as self-induced transparency
\cite{Maniadis2006} or supratransmission \cite{Geniet2002}.
For frequencies inside the linear magnetoinductive wave band the
power is transmitted along the array from the magnetoinductive wave modes, 
slightly modified by the presence of nonlinearity.  However, for frequencies 
in the pass-bands resulting from nonlinearity,
the power is transmitted along the array from nonlinear, breather-like excitations.
In the next Section we derive the model equations 
for a varactor-loaded SRR-based magnetic metamaterial, where the nonlinearity has 
the form a polynomial expansion on the main variable, like that 
investigated in \cite{Wang2008}. 
We also present bifurcation diagrams for the shortest possible end-driven array, 
i.e., an array with only two varactor-loaded SRRs,
from which only one is directly driven.
In Section III we obtain analytical solutions for the 
linear magnetoinductive waves of the end-driven, finite SRR array when losses are 
neglected, along with their discrete dispersion relation.
In Section IV we present numerical results for power transmission and discuss their
dependence on the model parameters for a short, end-driven varactor-loaded SRR array
with and without an absorbing boundary at the non-driven end.
We finish with concluding remarks in Section V.

\section{Varactor-Loaded SRR Array Model
}
Consider a ring-shaped or a squared split-ring resonator with a hyperabrupt tuning 
varactor mounted onto its slit. 
The varactor type could be selected from a large variety
of available varactors, according to the needs of the experiment. 
In recent experimental works on varactor-loaded SRR-based metamaterials
\cite{Wang2008,Poutrina2010,Larouche2010,Huang2010}, the varactor selected was 
a Skyworks SMV1231-079, whose voltage-dependent capacitance $C(U_D)$ 
($U_D$ is the voltage across the diode) is given by
\begin{equation}
  \label{1}
  C(U_D) = C_0 \left( 1 -\frac{U_D}{U_p} \right)^{-M} ,
\end{equation}
where $C_0$ is the DC rest capacitance, $U_p$ is the intrinsic potential, and 
$M$ is a parameter. The values of those parameters are provided by the 
manufacturer SPICE model for that varactor to be 
\begin{equation}
  \label{2}
    M=0.8,~~~ C_0 =2.2~pF,~~~ U_p =1.5 ~V .
\end{equation}
From Eq. (\ref{1}) we can determine the voltage-dependence of the normalized charge
\begin{equation}
  \label{111}
  q=\frac{Q_D}{C_0 U_p},
\end{equation}
with $Q_D$ being the charge in the diode, 
as a function of the normalized diode voltage $u=U_D/U_p$ as
\begin{equation}
  \label{3}
  q = \frac{1}{-M+1} \left[ 1 -\left( 1-u\right)^{-M+1} \right] .
\end{equation}
Assuming that $U_D < U_p$, the earlier equation can be solved for the diode voltage
as a function of the charge, i.e., $u=u(q)$
\begin{equation}
  \label{4}
  u = 1 - \left[ 1 -q\left(-M+1 \right) \right]^{\frac{1}{(-M+1)}} .
\end{equation}
When $|x| \equiv |q(-M+1)| << 1$, the voltage $u$ in the earlier equation can be expanded
in a Taylor series inpowers of $q$ around zero. Thus, neglecting term of order 
${\cal O}(q^4)$ or higher, we get 
\begin{equation}
  \label{5}
  u = q +\alpha q^2 +\beta q^3 ,
\end{equation}
where, by using Eq. (\ref{2}),  
\begin{equation}
  \label{5a}
   \alpha = -\frac{M}{2} =-0.4, \qquad \beta = \frac{1}{6} M (2M-1) =+0.08 .
\end{equation}
The expression Eq. (\ref{5}) can then be used in the voltage equation, obtained 
from Kirchhoff's voltage law, which represents the varactor-loaded SRR as an 
effective resistive-inductive-capacitive (RLC) circuit with voltage-dependent 
capacitance
\begin{equation}
  \label{6}
  L \frac{d^2 Q_D}{dt^2} +R \frac{dQ_D}{dt} +U_p u ={\cal E}(t) ,
\end{equation}
where 
\begin{equation}
  \label{222}
   U_p u \equiv U_D = U_p \left[
    \frac{Q_D}{C_0 U_p} +\alpha \left(\frac{Q_D}{C_0 U_p}\right)^2
                        +\beta \left(\frac{Q_D}{C_0 U_p}\right)^3 
    \right],
\end{equation},
${\cal E}(t) $ is the induced electromotive force of amplitude ${\cal E}_0$ 
and frequency $\omega$ 
resulting from an applied electromagnetic field with the same frequency, 
of the form
\begin{equation}
  \label{7}
  {\cal E}(t) = {\cal E}_0 \, \sin(\omega t) ,
\end{equation}
where $L$ is the inductance of the ring, $R$ is the sum of the Ohmic resistance
of the SRR ring and the series resistance of the diode, and $t$ the temporal variable.
The eigenfrequency $\omega_0$ of the circuit described by Eq. (\ref{6}) is given, 
in the linear limit without losses and driving, by
\begin{equation}
  \label{8}
  \omega_0 = \frac{1}{\sqrt{L\, C_0}} .
\end{equation}
That approximate model of the varactor-loaded SRR is found to be adequate
for varactor voltages not exceeding $0.5~V$ ($u \leq 1/3$ in normalized units), 
when the dissipative current of the varactor can be neglected \cite{Wang2008}.
%%%------ figure--1--------------------------------------------------------------
\begin{figure}[h!]
\includegraphics[angle=0, width=0.8\linewidth]{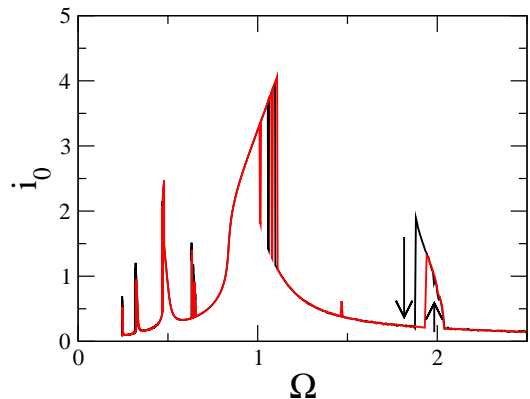}
\caption{
The current amplitude $i_0$ as a function of the normalized frequency 
$\Omega$ for a varactor-loaded SRR modeled by Eq. (\ref{10b}). 
Red curve: frequency increases; black curve: frequency decreases.
}
\end{figure}
%%%------ end-figure--1--------------------------------------------------------------

Using Eq. (\ref{111}) and the relations
\begin{eqnarray}
  \label{10a}
  \tau=t \, \omega_0, ~~  \Omega=\frac{\omega}{\omega_0}, 
  ~~ \varepsilon_0 =\frac{{\cal E}_0}{U_p} ,
\end{eqnarray}
Eq. (\ref{6}) can be written in the normalized form
\begin{eqnarray}
  \label{10b}
  \frac{d^2 q}{d\tau^2} +q +\alpha q^2 +\beta q^3 +\gamma \, \frac{d q}{d\tau} 
\nonumber \\
  =\varepsilon_0 \sin(\Omega \tau) ,
\end{eqnarray}
where 
\begin{equation}
  \label{120}
    \gamma= R\sqrt{C_0 /L}
\end{equation}
is the normalized loss coefficient.
The varactor-loaded SRR is a nonlinear oscillator, which exhibits multistability, 
hysteretic effects, and secondary resonances, when driven by a sufficiently strong 
external field.
A typical current amplitude $i_0$ - frequency $\Omega$ curve for the equivalent 
circuit model Eq. (\ref{10b}), like that shown in Fig. 1,
exhibits all those characteristics. 
Notice the hysteresis loop at normalized frequency $\Omega$ around $1.1$,
close to the resonance frequency of the linear system, and the very strong 
resonance at its second harmonic, i.e., at $\Omega \simeq 2$, where a smaller
hysteresis loop also appears.
Furthermore, we observe a subharmonic resonance at $\Omega \simeq 0.48$, 
as well as weaker resonances at $\Omega \sim 0.63$, $0.33$ and $\Omega \sim 0.25$.
For frequency intervals lying between the boundaries of the hysteresis loops
there are two different states of the oscillator,
with low and high current amplitude, which are simultaneously stable.
%%%------ figure--2--------------------------------------------------------------
\begin{figure}[h!]
\includegraphics[angle=0, width=0.85\linewidth]{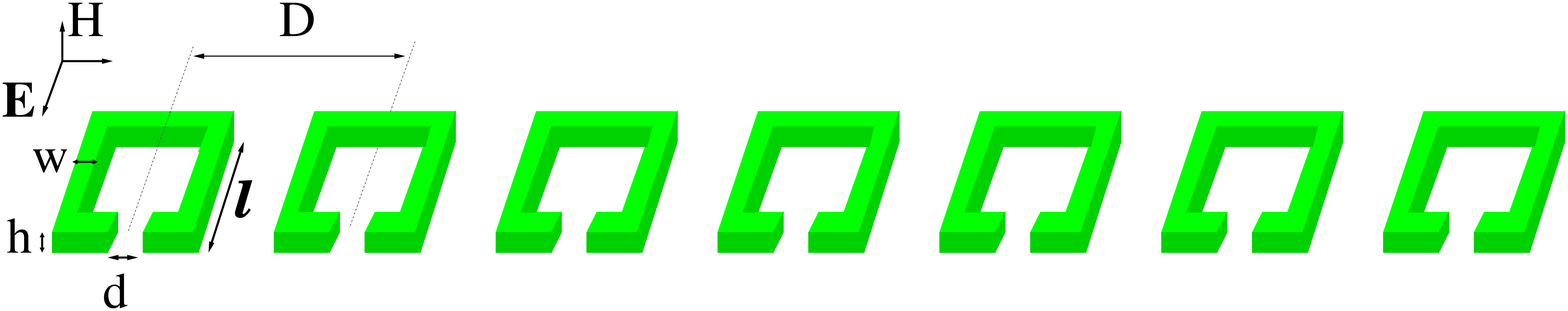} \\
\includegraphics[angle=-90, width=0.85\linewidth]{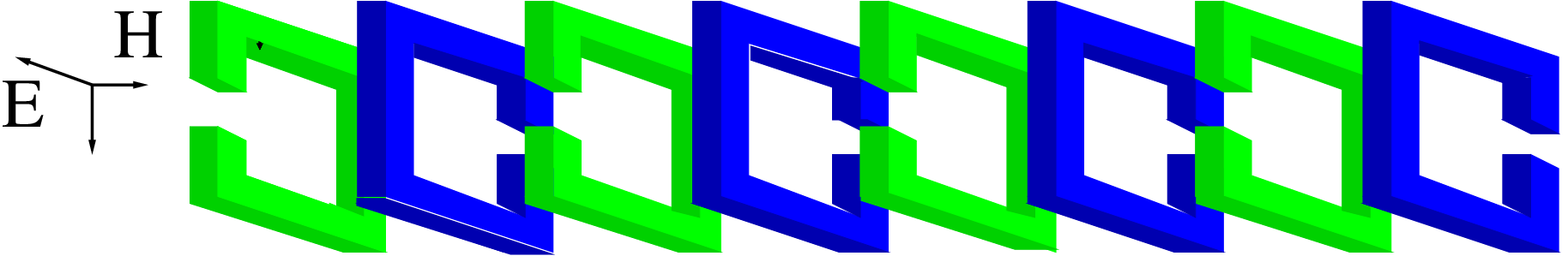}
\caption{
Schematic of an SRR array in the planar geometry (upper) and in the axial 
geometry (lower). 
}
\end{figure}
%%%------ end-figure--2--------------------------------------------------------------

Consider a one-dimensional periodic array of \emph{identical} varactor-loaded SRRs, 
which are coupled 
magnetically to their first neighbors through their mutual inductances. 
The array can be formed in two possible geometries, according to the relative
orientation of its elementary units (i.e., the SRRs) in the array.
As can be seen in Fig. 2, the SRRs can be arranged either in the planar geometry,
where all of them are lying on the same plane, or in the axial geometry,
where the axes of all the SRRs are lying on the same line.
The orientation of the incident electromagnetic field, 
which is also shown in the figure, is such that the magnetic component can excite 
an electromotive force in each SRR.  
Below we consider arrays that are driven only at one end (let us say the left end).
Thus only one SRR receives energy directly from the driver. However, the energy can 
be transmitted along the array due to the interaction between its elements. 
Including the coupling of each SRR to its nearest neighbors, we get for the 
charge $Q_{D,n}$ of the $n-$th varactor the dynamic equations
\begin{eqnarray}
  \label{9}
   L \frac{dI_n}{dt} +R I_n + U_p u_n +M \frac{dI_{n-1}}{dt} 
  +M\frac{dI_{n+1}}{dt} = {\cal E} \delta_{n,1} ,
\end{eqnarray}    
where $n=1,...,N$, with $N$ being the total number of SRRs in the array,
$I_n (t) = {d Q_{D,n}}/{dt}$ is the current flowing in the $n-$th SRR,
$U_p u_n$ is the approximate voltage across the $n-$th varactor, and 
$M$ (not to be confused with the varactor parameter) is the mutual inductance 
between two neighboring SRRs. 
Using the relations from Eq. (\ref{10a}), the coupled nonlinear Eqs. (\ref{9}) 
can be written in normalized form as      
\begin{eqnarray}
  \label{11}
  \frac{d^2}{d\tau^2} \left( \lambda \, q_{n-1} +q_n + \lambda \, q_{n+1}
                \right) 
    + q_n +\alpha q_n^2 +\beta q_n^3 =
\nonumber \\
-\gamma \, \frac{d q_n}{d\tau}  +\varepsilon_0 \sin(\Omega \tau) \delta_{n,1} ,
\end{eqnarray}
where $\lambda= M/L$ is the coupling coefficient between neighboring SRRs. 
The function $\delta_{n,1}$ is 
unity for $n=1$, while it is zero for any other $n$. 
The coupling parameter $\lambda$ may assume both positive and negative values,
corresponding to planar and axial geometry of the varactor-loaded SRR array,
respectively.

For a particular array, the values of the coupling and loss coefficients $\lambda$ 
and $\gamma$, respectively, can be estimated from its geometrical and material 
parameters. Specifically, the mutual inductance $M$, the ring inductance $L$, and 
the Ohmic resistance $R_{SRR}$ of the ring can be calculated from the following expressions
\begin{eqnarray}
  \label{13}
    M&=&\mu_0 \frac{\pi a^2 a^2}{4 D^3}, \\
  \label{14}
    L&=&\mu_0 a \left[ \ln\left( \frac{16 a}{h} \right) -1.75 \right], \\
  \label{15}
    R_{SRR}&=&\frac{2 a}{\sigma h \delta} ,
\end{eqnarray}
where 
$\mu_0$ is the magnetic permeability in vacuum, 
$a$ is the average radius of each SRR,
$h$ the diameter of the wire of each SRR,
$D$ is the center to center distance between neighboring SRRs, and
$\sigma$ and $\delta$ is the conductivity and skin depth of the SRR wire.
Eqs. (\ref{13})--(\ref{15}) refer to circular SRRs with circular cross-section.
However, they can also be used for circular SRRs with square cross-section  
by defining an equivalent diameter of the SRR wire $h=\sqrt{4 d w /\pi}$, 
with $d$ and $w$ being the depth and width, respectively, of the SRR wire.
For $a\simeq 3.7~mm$, $D\simeq 11~mm$, $h\simeq 0.5~mm$, we get from 
Eqs. (\ref{13}) and (\ref{14}), respectively, that $M\simeq 0.14~nH$ and $L\simeq 14~nH$,
and consequently $\lambda \simeq 0.01$ .
For $\delta \simeq 2.2\times 10^{-3} ~mm$ and  
$\sigma \simeq 0.5\times 10^6 ~(\Omega cm)^{-1}$
(for an SRR made of copper wire at $\sim 1~GHz$ at room temperature) 
we get from Eq. (\ref{15}) that $R_{SRR} \simeq 0.13 ~\Omega$.
The diode series resistance $R_s$ of the varactor used in Ref. \cite{Wang2008}
is around $\simeq 2~\Omega$; that is, most of the Ohmic resistance of the varactor-loaded
SRR, $R=R_{SRR} + R_s =0.13 +2.00 =2.13~\Omega$ comes from the varactor.
By substituting $R \simeq 2.13~\Omega$ in the second of Eq. (\ref{120}), along with
$L=14~nH$ and $C_0 =2.2~pF$, we get that $\gamma =0.027$.
The resonance frequency of a single varactor-loaded SRR is 
$f_0 \simeq (2\pi \sqrt{L \, C_0})^{-1} \simeq 0.9~GHz$.      
In the numerical calculations below we used the estimated value of $\lambda$ 
that indicates weakly coupled SRRs.
However, we use a smaller value of $\gamma$, which could be achieved in practice
by using a metalic wire with higher conductivity and an ultra-low resistance varactor
with similar capacitance-voltage relation.
%%%------ figure--3--------------------------------------------------------------
\begin{figure}[t!]
\includegraphics[angle=0, width=0.85\linewidth]{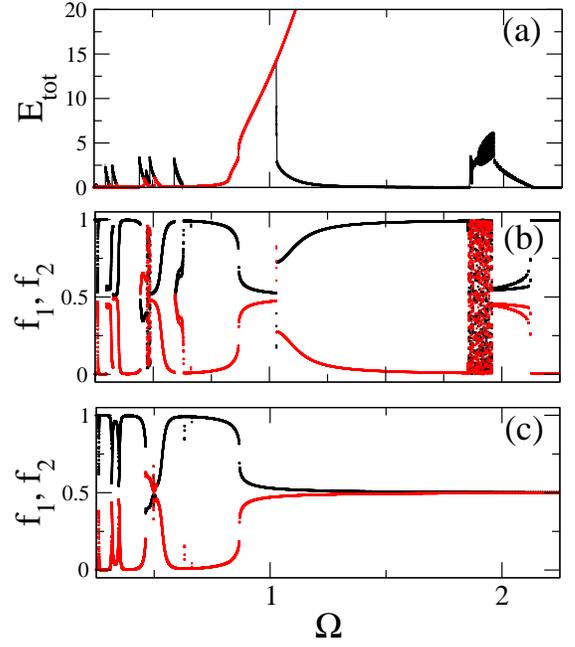}
\caption{
(a) Energy $E_{tot}$ bifurcation diagram for varying frequency $\Omega$ for 
an asymetrically driven varactor-loaded SRR dimer with $\gamma=0.001$,
$\varepsilon_0 =0.3$, and $\lambda=+0.01$.
(b) Energy fractions $f_1$ and $f_2$ 
of the first (driven) and the second SRR, respectively, of the dimer
in the low energy state.
(c) Energy fractions $f_1$ and $f_2$ 
of the first (driven) and the second SRR, respectively, of the dimer
in the high energy state.
}
\end{figure}
%%%------ end-figure--3--------------------------------------------------------------

The shortest end-driven varactor-loaded SRR array is that comprised of two 
elements. Since only one of them is driven by the external field, we refer
to this system as the asymmetrically driven nonlinear dimer. The increase of 
degrees of freedom leads to interesting and complicated behavior, 
that is marked by the appearance of chaos.
In Fig. 3, the bifurcation diagram of the total energy $E_{tot}$ of the dimer 
with varying normalized frequency $\Omega$ is shown for positive coupling $\lambda$.
For negative coupling $\lambda$ with equal strength we get very similar results
that share a number of characteristic features.
First of all, multistability is again present, for normalized frequencies 
in an interval around $\Omega \sim 0.9$.
There, two simultaneously stable energy states, with high and low energy, co-exist,
as it is shown in Fig. 3a.
The energy difference between those states increases condiderably with increasing 
frequency. 
Moreover, low energy state becomes chaotic for frequencies around $\Omega \sim 1.9$. 
It is also interesting to see how the total energy is divided between the two elements.
The energy fractions $f_1=E_1/E_{tot}$ and $f_2 = E_2/E_{tot}$,
with $E_1$ and $E_2$ being the energies of the first (driven) and second SRR
of the dimer, are shown both for the high and the low energy states in Fig. 3b and 3c
as a function of the normalized frequency $\Omega$. When the dimer is in the low 
energy state, most of the energy is concentrated in driven SRR whose energy fraction
is close to unity (correspondingly, the energy fraction of the other SRR is close to zero).
However, for frequencies around the resonances, energy can be transfered easily from 
the one SRR to the other, and then the energy fractions of the SRRs in the dimer
attain comparable values. In the chaotic regions, the energy is tranfered irregularly
from one SRR to the other and the energy fractions span the whole interval between 
zero and one. 
In the high energy state (Fig. 3c), the energy fractions are almost equal for 
frequencies above the linear resonance frequency, $\Omega \sim 1$. 
Below that frequency, the energy is either concentrated in the driven SRR 
(for frequencies far from resonances), or the values of the energy fractions
are comparable (close to resonances.

\section{Driven Linear Solutions 
}
For a weakly driven array, the nonlinear terms which are proportional to 
$\alpha$ and $\beta$ are not so important and they can be neglected.
Then, the system of Eqs. (\ref{11}) reduces to a linear one
\begin{eqnarray}
  \label{16}
  \frac{d^2}{d\tau^2} \left( \lambda \, q_{n-1} - q_n + \lambda \, q_{n+1}
                \right) 
    + q_n 
  \nonumber \\
  =-\gamma \, \frac{d q_n}{d\tau}  
   +\varepsilon_0 \sin(\Omega \tau) \delta_{n,1} .
\end{eqnarray}
The dispersion
relation of the linear system can be obtained by the substitution
$q_n=A \cos(\kappa n -\Omega\tau)$ in the absence of losses and applied field,
i.e., for $\varepsilon_0 =0$ and $\gamma=0$, which gives
the linear dispersion of magnetoiductive waves 
\begin{equation}
\label{17}
  \Omega_{\bf \kappa} = \frac{1}{\sqrt{1 +2\, \lambda \, \cos(\kappa)}} ,
\end{equation}
where $\kappa$ is the (normalized) wavevector ($-\pi \leq \kappa \leq \pi$). 
Eq. (\ref{17}) defines a frequency band of width $\Delta\Omega \simeq 2\lambda$
that is bounded by a minimum and a maximum frequency
$\Omega_{min} = 1/\sqrt{1+2\lambda}$ and $\Omega_{max} = 1/\sqrt{1-2\lambda}$,
respectively.

However, it is also possible to calculate the dispersion, expressed as a series 
of resonant frequencies, and the exact form of the linear magnetoinductive modes
for the finite system in the absence of losses ($\gamma=0$). 
For finite systems, Eq. (\ref{16}) should be implemented with 
free-end boundary conditions, to account for the termination of the structure, i.e., 
\begin{equation}
  \label{17a}
    q_0 = q_{N+1} = 0 .
\end{equation}    
By substituting $q_n = Q_n \, \sin(\Omega \tau)$ into Eq. (\ref{16}), the stationary 
equations can be written as
\begin{equation}
\label{18}
  s Q_{n-1} + Q_n + s Q{n+1} = \kappa ,
\end{equation}
where 
\begin{equation}
\label{19}
  s=-\frac{\lambda \Omega^2}{1 -\Omega^2}, \qquad
  \kappa =\frac{\varepsilon_0}{1 -\Omega^2} \delta_{n,1} ,
\end{equation} 
or, in matrix form
\begin{equation}
  \label{20}
    {\bf Q} = \kappa\, \hat{\bf S}^{-1} {\bf E}_1 ,
\end{equation}
where  ${\bf Q}$ and ${\bf E}_1$ are $N-$dimensional vectors  with
componets $Q_i$ and $\delta_{i,1}$, respectively,
and $\hat{\bf S}^{-1}$ is the inverse of the $N\times N$  coupling matrix
$\hat{\bf S}$. 
The latter
is a real, symmetric tridiagonal matrix that has diagonal elements 
equal to unity, while all the other non-zero elements are equal to $s$.
The need to find the inverse of tridiagonal matrices like $\hat{\bf S}$
arises in many scientific and engineering applications.
Recently,  Huang and McColl \cite{Huang1997}
related the inversion of a general trigiagonal matrix to second order 
linear recurrences, and they provided a set of very simple analytical formulae
for the elements of the inverse matrix.
Those formulae lead immediately to closed forms for certain trigiagonal matrices,
like $\hat{\bf S}$ \cite{Lazarides2010a}.
The components of the ${\bf Q}$ vector can be written as
\begin{equation}
  \label{21}
   Q_i = \kappa \left( \hat{\bf S}^{-1} \right)_{i1} ,
\end{equation}
where $\left( \hat{\bf S}^{-1} \right)_{i1}$ is the $(i1)-$element of the inverse
coupling matrix $\hat{\bf S}^{-1}$, whose explicit form is given in 
\cite{Lazarides2010a}. Then, the solution of the linear system 
Eq. (\ref{16}) and (\ref{17a}) for any driving frequency and finite $N$ is given by 
\begin{eqnarray}
  \label{22} 
    q_n (\tau) = \kappa \mu \frac{\sin[(N-n+1)\theta']}{\sin[(N+1)\theta']}
       \sin(\Omega \tau) ,
\nonumber \\
  \theta' = \cos^{-1} \left( \frac{1}{2|s|} \right) , 
\end{eqnarray}
for $s>+1/2$ and $s<-1/2$, 
and 
\begin{eqnarray}
  \label{23} 
    q_n (\tau) = \kappa \mu \frac{\sinh[(N-n+1)\theta]}{\sinh[(N+1)\theta]} 
    \sin(\Omega \tau) ,
\nonumber \\
  \theta = \ln\frac{1+\sqrt{1-(2s)^2}}{2|s|} , 
\end{eqnarray}
for $-1/2 < s < +1/2$,
where
\begin{eqnarray}
  \label{24}
   \mu= \frac{1}{|s|} \left( -\frac{|s|}{s} \right)^{n-1} .
\end{eqnarray} 
Note that the $Q_i$s are uniquely determined by the parameters of the system.
%%%------ figure--4--------------------------------------------------------------
\begin{figure}[h!]
\includegraphics[angle=0, width=0.85\linewidth]{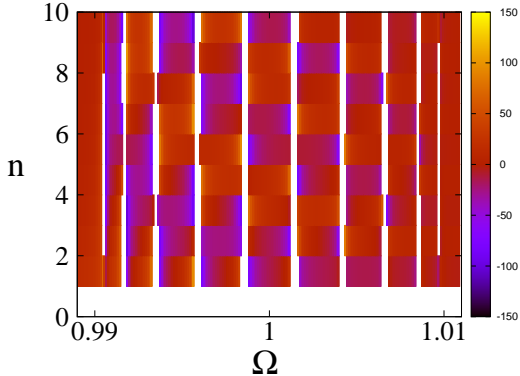}
\caption{
Density plots of the $Q_n$'s in a stationary state, 
on the site number $n$ - normalized frequency $\Omega$ plane 
calculated from the analytical expressions Eqs. (\ref{22}) and (\ref{23})
for $\lambda=+0.01$ and $\varepsilon_0 =0.3$. 
The vertical white bands indicate the positions of the resonances. 
}
\end{figure}
%%%------ end-figure--4--------------------------------------------------------------

From the analytical solution in the linear magentoinductive wave band, Eq. (\ref{22}),
which corresponds to either $s>+1/2$ or $s<-1/2$,
we can infer the resonant frequencies from zeroing the denominator, 
i.e., by setting $\sin[(N+1)\theta'] =0$. Then we see that the solution has a 
resonance for 
\begin{eqnarray}
  \label{25}
  s\equiv s_m=\frac{1}{2 \cos\left[ \frac{m\pi}{(N+1)} \right]} ,
\end{eqnarray}
where $m$ is an integer ($m=1,...,N$). The resonant freqquencies are obtained
by solving the first of Eq. (\ref{19}) with respect to $\Omega$, and substituting 
the resonant values of $s\equiv s_m$ from Eq. (\ref{25}). 
Then we get the discrete dynamic
dispersion relation of linear magnetoinductive waves in the varactor-loaded SRR array, 
the discrete analogue of Eq. (\ref{17}), as 
\begin{eqnarray}
  \label{26}
   \Omega_m =\frac{1}{\sqrt{ 1 -2\, \lambda \, \cos\left( \frac{m\pi}{N+1} \right) }}  ,
\end{eqnarray}
where $m$ is the mode number ($m=1,...,N$). 
In Fig. 4, a density plot of the $Q_n$'s on the $n-\Omega$ plane, 
calculated analytically from Eqs. (\ref{22}) and (\ref{23}),
is shown for positive coupling coefficient $\lambda$.
In the frequency interval shown, the $Q_n$'s 
have rather large values because, due to the narrow band, all frequencies 
are very close to a resonance. 
Moreover, the analytical solutions presented above have been obtained
in the absence of losses, so that there is nothing to prevent the  $Q_n$'s
from increasing indefinitely close to a resonance (and to go to infinity
exactly at resonance). The white vertical regions in these figures correspond
to frequency intervals around a resonance, where the  $Q_n$'s have very large values.
Thus, those regions indicate the frequencies of the resonances.
We can also see in Fig. 4 that the solutions close to the lower bound 
($\Omega_{min} \simeq 0.99$) of the linear magnetoinductive wave band exhibit 
very high amplitudes. For negative coupling coefficient with equal strength
the high amplitude solutions appear close to the upper bound 
($\Omega_{max} \simeq 1.01$) of the linear magnetoinductive wave band. 

%%%------ figure--5--------------------------------------------------------------
\begin{figure}[t!]
\includegraphics[angle=0, width=0.85\linewidth]{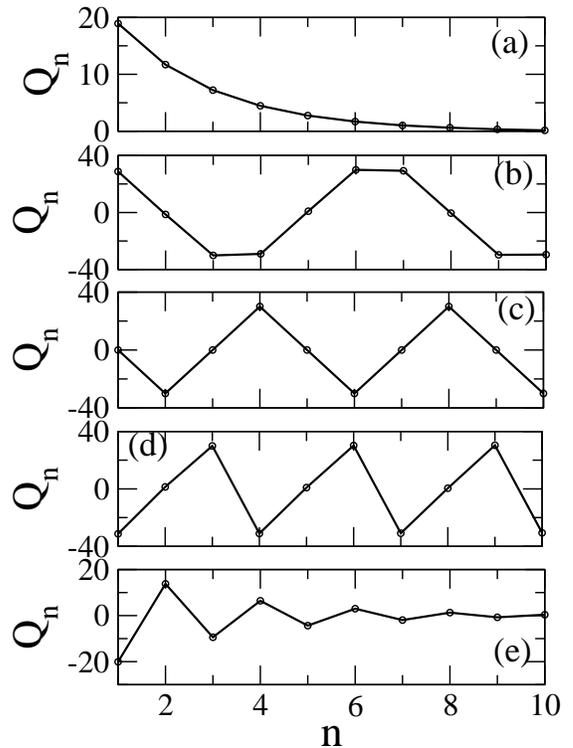}
\caption{
Typical profiles calculated analytically from Eqs. (\ref{22}) and (\ref{23})
for $\lambda=+0.01$, $\varepsilon_0 =0.3$, and five different frequencies: 
(a) $\Omega=0.989$; (b) $\Omega=0.995$;(c) $\Omega=1.000$;
(d) $\Omega=1.005$; (e) $\Omega=1.011$.
The frequencies in (a) and (e) lie outside the linear magnetoinductive 
wave band.
}
\end{figure}
%%%------ end-figure--5--------------------------------------------------------------
In Fig. 5 the profiles of the analytical solutions for five different frequencies,
both inside and outside the linear magnetoinductive wave band are shown for 
positive $\lambda$. The frequencies in Figs. 5a and 5e lie outside the linear band,
so that the $Q_n$'s decrease exponentially with increasing site number $n$.

\section{Power transmission along the array.
}
In the following we focus on the power transmission along a varactor-loaded SRR 
array with $N=10$ elements, which is driven at the left end ($n=1$). 
The average power dissipated in the $n-$th SRR is defined as
\begin{eqnarray}
  \label{27}  
    P_n = R < I_n^2 (\tau) >_{\tau_0} ,
\end{eqnarray}
where $< >_{\tau_0}$ denotes time-average over $\tau_0$ time-units.
The power density $P_n$ can be normalized to $P_0= I_0^2 R$,
where $I_0 =\omega_0 \, C_0 \, U_p$ and $R$ the value of resistance
that results from the value of the dissipation coefficient used in 
the simulations. For $\gamma=0.001$ we have that $R\simeq 0.06$
and $P_0 \simeq 28~\mu W$. 
It is convenient to express $P_n$ in dBm, according to the relation
\begin{eqnarray}
  \label{28}
   P_n (dBm) = 20 \log\left( <i_n^2>_{\tau_0} \frac{P_0 (in~W)}{1~mW} \right)
\nonumber \\
\simeq 4.34 \ln( 0.028 < i_n^2 >_{\tau_0} ) ,
\end{eqnarray}
where $i_n$ is the normalized current in the $n-$th SRR. 

%%%------ figure--6--------------------------------------------------------------
\begin{figure}[h!]
\includegraphics[angle=0, width=0.85\linewidth]{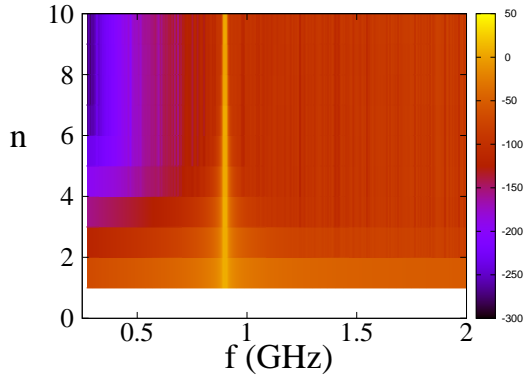} \\ 
\caption{ 
Average power density plot (in dBm) on the $n - f$ plane, 
for $\lambda =+0.01$, $\gamma=0.001$, and $\epsilon_0 = 0.3$,
in the linear regime ($\alpha=0$, $\beta=0$). 
}
\end{figure}
%%%------ end-figure--6--------------------------------------------------------------
The system of Eqs. (\ref{11}) is integrated with a 4th order Runge-Kutta
algorithm with fixed time-stepping, and  boundary conditions
$q_0 = q_{N+1} =0$ to account for the termination of the structure. 
The initial conditions were set to zero.
The time allowed for the elimination of transients in each value of the frequency
is $2000 ~T_0$ ( $T_0 = 2\pi/\omega_0$), while the time interval $\tau_0$ over which 
the average  power is calculated is typically $1000 ~T_0$. In order to convert
time intervals to  physical units one has to multiply by the inverse of the 
resonance frequency $f_0^{-1}$, which gives $\tau_0 \simeq 2.2~\mu s$.
In the linear regime, i.e., $\alpha=0$ and $\beta =0$, 
the average power $P(dBm)$ density plot on the site number $n$ - frequency $f$ 
plane shown in Fig. 6 exhibits the expected behavior;
that appreciable power transmission only occurs for frequencies in the linear band,
i.e., around $f=0.9~GHz$. 
%%%------ figure--7--------------------------------------------------------------
\begin{figure}[h!]
\includegraphics[angle=0, width=0.85\linewidth]{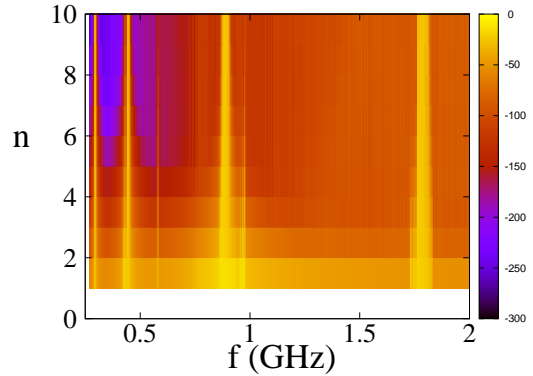} \\ 
\caption{ 
Average power density plot (in dBm) on the $n - f$ plane, for $\lambda =+0.01$,
$\gamma=0.001$, $\alpha=-0.4$, $\beta=0.08$, and
$\epsilon_0 = 0.3$.
%%; $\epsilon_0 = 0.29$ (middle); $\epsilon_0 = 0.28$
}
\end{figure}
%%%------ end-figure--7--------------------------------------------------------------
However, the activation of the nonlinear terms reveals that even a relatively 
low driving amplitude $\varepsilon_0=0.3$ is sufficient for nonlinear power
transmission bands to bew formed. Thus, power can be transmitted not only for 
frequencies in the linear band, but also in the otherwise forbidden frequency
regions. As it can be observed in Fig. 7, a number of transmission bands have 
appeared due to the nonlinearity. The array thus becomes transparent in power
transmission at frequency
intervals around the nonlinear resonances of a single varactor-loaded SRR 
oscillator. That self-induced transparency is a robust effect and survives
with considerably increased losses, as it is shown in Fig. 8, where the loss 
coefficient has been increases by an order of magnitude (from $0.001$ to $0.01$.
In the density plots shown above, significant power transmission is indicated by 
red color. In both Figs. 7 and 8, nonlinear power transmission occurs at
$f \sim 0.3~GHz$, $f \sim 0.45~GHz$, and $f \sim 1.8~GHz$. 
The latter nonlinear band, around the second harmonic of the resonance frequency 
of the linear system, seems to be the wider of all.
%%%------ figure--8--------------------------------------------------------------
\begin{figure}[h!]
\includegraphics[angle=0, width=0.85\linewidth]{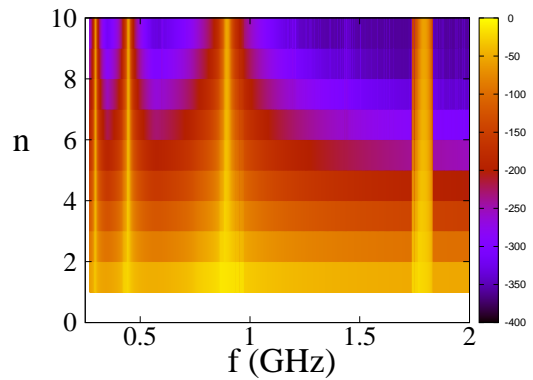} \\ 
\caption{ 
Average power density plot (in dBm) on the $n - f$ plane, for $\lambda =+0.01$,
$\gamma=0.01$, $\alpha=-0.4$, $\beta=0.08$, and
$\epsilon_0 = 0.3$.
%% (upper); $\epsilon_0 = 0.29$ (middle); $\epsilon_0 = 0.28$
}
\end{figure}
%%%------ end-figure--8--------------------------------------------------------------

When losses are small, the appearance of more nonlinear bands is possible,
like the one seen just above the linear band in Fig. 7, at $f \sim 1~GHz$.
That transmission band is due to an interplay between nonlinearity and geometrical
(lattice) resonances, and it is very sensitive to changes in the parameters
of the model. 
Note that in the density plots presented above the frequency is given in natural 
units, using the relation $f=f_0 \, \Omega =0.9 \Omega~GHz$.

It is very illuminating to see the time evolution of the current flowing in 
each SRR of the array for some frequencies in the transmission bands. We have chosen 
two different frequencies, one in the linear band and one the high frequency band 
which is around the second harmonic of the linear resonance frequency.
The time evolution of the currents in the first, fourth, seventh, and tenth (last)
SRR are shown for $f=0.888~GHz$ and $f=1.78~GHz$ in Figs. 9a and 9b, respectively.
The currents oscillate periodically in all SRRs, although with different amplitudes
and relative phases with respect to the driver. This figure suggests that power
transmission for frequencies in the forbidden band gap occurs through the linear
modes of the system. For frequencies outside the linear and nonlinear bands, 
the currents exhibit similar oscillations, although their amplitude becomes vanishingly
small with increasing site number $n$.  
%%%------ figure--9--------------------------------------------------------------
\begin{figure}[h!]
\includegraphics[angle=0, width=0.8\linewidth]{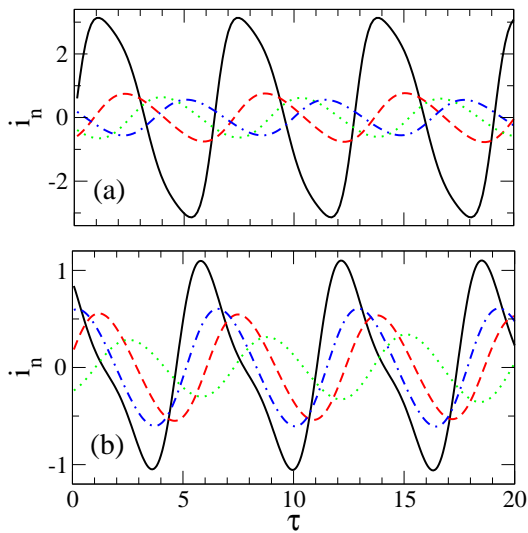}
\caption{ 
Time dependence of the current $i_n$ in the $n-$th SRR for an array 
with $N=10$, $\lambda = 0.01$, $\epsilon_0 = 0.3$, $\gamma=0.001$,
and frequency (a) $f=0.888~GHz$; (b)  $f=1.78~GHz$. 
The four different curves are for the current in the $n=1$ SRR (black-solid),
$n=4$ SRR (red-dashed), $n=7$ SRR (green-dotted), $n=10$ SRR (blue-dot-dashed).   
}
\end{figure}
%%%------ end-figure--9--------------------------------------------------------------
The relative efficiency of power transmission in the two frequencies 
used in the previous figure, are shown in Fig. 10, along with the power 
transmission for a specific frequency ($f=1.197~GHz$) outside all linear and 
nonlinear bands. It is remarkable that the transmission efficiency is
almost the same for frequencies in the two pass-bands, while for the other
one the transmitted power practically vanish at the fourth site. 
%%%------ figure--10--------------------------------------------------------------
\begin{figure}[h!]
\includegraphics[angle=0, width=0.8\linewidth]{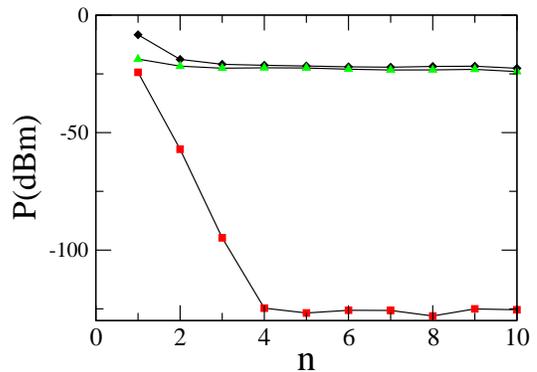} 
\caption{ 
Power tranfer efficiency in different frequency regions, i.e.,
average power (in dBm) as a function of site number $n$ for an array
$\varepsilon_0 = 0.3$, $\lambda=0.01$, $\gamma=0.001$, $N=10$, and
$f=0.888~GHz$ (black diamonds); $f=1.197~GHz$ (red squares), 
$f=1.78~GHz$ (green triangles).
}
\end{figure}
%%%------ end-figure--10--------------------------------------------------------------

In the density plots above, the right boundary was actually a reflecting one,
allowing the formation of stationary states in the array.
We have also used a totally absorbing boundary by adding five more sites at the 
right end of the array, where the damping coefficient increases exponentially
with increasing site number. Thus, the waves coming from the left are completely
damped at the end of the extended array, and no reflection occurs.
The total number of SRRs is now $N'=N +N_1$ where $N_1=5$, and the dissipation 
coefficient $\gamma$ is now a function of the site number $n$, $\gamma(n)$, given by 
\begin{equation}
\label{30}
  \gamma (n) =
\left\{
\begin{array}{cl}
\gamma & , \ 1 \leq n \leq N \\
\gamma e^{\mu n}  & , \ N < n \leq N'
\end{array}
\right. ,
\end{equation}
where $\mu = \frac{ \log(1/\gamma) }{N_1}$, with $\gamma=0.001$ as before.
For the array with $n-$dependent dissipation coefficient, 
a typical average power density plot is shown in Fig. 11, where the power
in each site is the average over a time interval of $5000~T_0$.
We observe, at least for the first $N$ sites of the array, a pattern similar with 
those seen in Figs. 7 and 8. That is, the appearance of the linear and the 
nonlinear transmission bands at about the same frequency intervals.
We also observe that almost all of the transmitted power is absorbed from 
the five last sites of the array. Importantly, the narrow band coming from 
geometrical resonances in Figs. 7 and 8, which is around  $f=1~GHz$,
appears in Fig. 11 as well, and moreover it seems that it is capable of 
transmitting power to the end of the array. 
%%%------ figure--11--------------------------------------------------------------
\begin{figure}[h!]
\includegraphics[angle=0, width=0.85\linewidth]{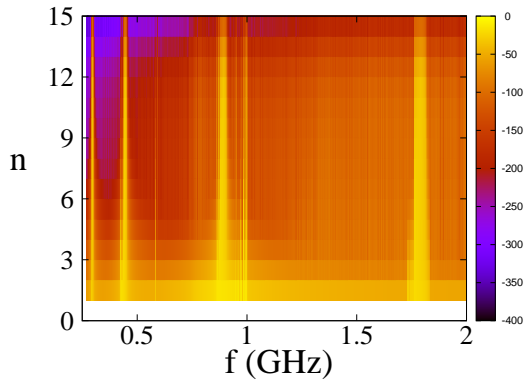} \\ 
\caption{ 
Average power density plot $P_n$ (in dBm) on the $n - f$ plane for a 
varactor-loaded SRR array 
with absorbing right boundary (see text), for  
$\lambda =+0.01$, $\epsilon_0 =0.297$, $\gamma=0.001$, and $N'=15$ ($N=10$).
}
\end{figure}
%%%------ end-figure--11--------------------------------------------------------------

\section{Discussion and Concluding Remarks
}
We have used a simple electric circuit model of inductively coupled
resistive-inductive-capacitive circuits with voltage-dependent
capacitance to investigate the transmission of power in a 
one-dimensional, end-driven, nonlinear magnetic metamaterial comprised 
of varactor-loaded SRRs.  
We have used both a reflecting and an absorbing boundary for the end that 
is not driven by the external field.
The importance of such discrete models for describing the dynamic 
behaviour of SRR-based magnetic metamaterials has been recently 
appreciated \cite{Liu2009}, although similar discrete models 
have been employed earlier for the study of localized excitations in 
those systems \cite{Shadrivov2006b,Lazarides2006}
  
In the present work, we have focused on the transmission of power along
the array, and especially that in the nonlinear regime, 
which results in the formation of nonlinear pass-bands for frequencies
close to the resonances of a single varactor-loaded SRR that is described 
by the approximate model equation (\ref{10b}). 
The approximate model holds fairly well for relatively low driving amplitudes,
for which the dissipative diode current of the varactor can be neglected.
It seems, thus that the 
properties of the individual elements of the array predominantly determine
its transmission properties, at least in the case of weak magnetoinductive 
coupling investigated in the present work. 
The numerical results demonstrate that the nonlinear pass-bands transmit
power with efficiently comparable to that of the linear band. 
For the parameter sets used in this work, which are close to those 
in \cite{Wang2008} for a single varactor-loaded SRR, there are three nonlinear
bands, both above and below the linear band.
Their width apparently depends on the frequency, i.e., the bands at higher 
frequencies are significanlty wider than those at lower frequencies.
 
Moreover, we have also observed the formation, either complete or partial,
of another nonlinear pass-band at frequencies slightly above the linear
band (at $f \sim 1~GHz$). That band, which is rather narrow and can be 
observed both in Figs. 7 and 8 as well as in Fig. 11,
seem to depend sensitively on the model parameters 
$\lambda$, $f$, $\varepsilon_0$ (for $\alpha$ and $\beta$ fixed),
and mainly on the loss coefficient $\gamma$. 
Comparing Figs. 7 and 8, which have been obtained for different loss 
coefficient $\gamma=0.001$ and $0.01$, respectively, we see that 
the power transmission in that band considerably decreases when the losses
increase by an order of magnitude.
That type of nonlinear band appears due to geometrical resonances
of the standing waves that are formed in the array due to driving at one end.
Geometrical resonances seem to offer an important mechanism for the generation
of nonlinear pass-bands,  
which can be employed for constructing transmission lines with banded power 
transmission spectra with bands at desired frequency ranges.
Similar conduct-free transmission lines could be formed by superconducting 
magnetic metamaterials, where the basic structural element
(the SRR) is replaced by its direct superconducting analogue, the rf SQUID
\cite{Lazarides2007,Lazarides2008b}. The rf SQUID, where the acronym stands 
for radio-frequency superconductive quantum interference device, 
is just a superconducting ring interrupted by a Josephson junction,
which makes it an intrinsically nonlinear element.
A one-dimensional array of rf SQUIDs coupled through their mutual inductances
forms a nonlinear magnetoinductive transmission line \cite{Lazarides2008b}
whose power transmission properties is a matter of future work.

\section*{Acknowledgments} 
\noindent 
This work was supported in part by the European Office of Aerospace Research 
and Development, AFSOR award FA8655-10-1-3039,
and by the EURYI and MEXT-CT-2006-039047 .

%\end{multicols}

\end{document}